# Wireless Communications in Doubly Selective Channels with Domain Adaptivity

J. Andrew Zhang, Hongyang Zhang, Kai Wu, Xiaojing Huang, Jinhong Yuan, Y. Jay Guo

*Abstract*: **Wireless communications are significantly impacted by the propagation environment, particularly in doubly selective channels with variations in both time and frequency domains. Orthogonal Time Frequency Space (OTFS) modulation has emerged as a promising solution; however, its high equalization complexity, if performed in the delay-Doppler domain, limits its universal application. This article explores domain-adaptive system design, with an emphasis on adaptive equalization, while also discussing modulation and pilot placement strategies. It investigates the dynamic selection of best-fit domains based on channel conditions to enhance performance across diverse environments. We examine channel domain connections, signal designs, and equalization techniques with domain adaptivity, and highlight future research opportunities.**

*Index Terms*: **Doubly selective channels, OTFS, OFDM, Domain adaptivity.**

## I. INTRODUCTION

The performance of wireless communication systems is significantly affected by the propagation environment. In particular, doubly selective channels, characterized by variations in both time and frequency, present a formidable challenge. These channels exhibit different appearances in different domains, such as time, frequency, Doppler, and delay, necessitating sophisticated modulation and equalization techniques to maintain reliable communications.

Broadband wireless communications mainly face slow-time-varying frequency selective channels where multipath delay spread is large. For such channels, conventional single carrier (SC), where data symbol modulation and equalization are both performed in the time domain, becomes inefficient due to the high complexity of equalization. Outperforming SC, Orthogonal Frequency Division Multiplexing (OFDM) has become the de-facto modulation, for its effectiveness in equalizing frequency-selective fading channels. It also enables efficient resource allocation and optimization due to the multiplicative relationship between the signal and channel. Moreover, precoded OFDM, such as discrete Fourier transform (DFT)-precoded OFDM in 5G mobile networks, has been widely used to improve frequency diversity and/or reduce peak-to-average power ratio (PAPR) of OFDM signals. However, conventional OFDM struggles in *doubly selective channels* with large Doppler shifts and rapid time variations, typically encountered in high-mobility applications.

J.A. Zhang (*Senior member, IEEE*), H. Zhang, K. Wu, X. Huang (*Senior member, IEEE*), and Y.J. Guo (*Fellow, IEEE*) are with the University of Technology Sydney, Australia. Email: {Andrew.Zhang; Hongyang.Zhang; Kai.Wu; Xiaojing.Huang; Jay.Guo}@uts.edu.au.
J. Yuan (*Fellow, IEEE*) is with the University of New South Wales, Australia. Email: J.Yuan@unsw.edu.au.

To address the limitations of OFDM in doubly selective channels, Orthogonal Time Frequency Space (OTFS) modulation [1,2] and its variants, such as Orthogonal Delay-Doppler Division Multiplexing [3] and Zak-OTFS [4], have been proposed. OTFS modulates data symbols in the delay-Doppler (dD-) domain, offering improved resilience against time variations. Research has demonstrated that OTFS can significantly enhance performance in high-mobility scenarios. However, the equalization complexity of OTFS in the dD-domain may dramatically increase and performance degrades, as the channel sparsity decreases [1,2]. This makes conventional delay-Doppler domain equalization *not* a practically universal solution for all types of doubly selective channels. Recall that OFDM with frequency-domain equalization is preferable in dense multipath channels due to the high complexity of time-domain equalization. It is straightforward to see that equalization in the delay-Doppler domain may not be ideal for time-varying channels with large delay spread. Equalization for OTFS in alternative domains has been investigated in, e.g., [5-8], demonstrating the necessity of domain adaptive designs.

This article explores signal design and equalization that leverage domain adaptivity for wireless communications in doubly selective channels. Introducing the viewpoint of precoding, data modulation, pilot and channel estimation, and equalization can be implemented in different domains. In addition, dynamically adapting the equalization domain -- whether time, frequency, Doppler, or delay -- based on real-time channel conditions can lead to optimized performance with low complexity. This article unveils this overlooked potential by identifying best-fit domains for different channel conditions, enhancing signal processing performance and complexity across diverse propagation environments.

In the rest of this article, we delve into the technical details of this approach. The primary focus of this article is to highlight the significance of domain-adaptive design and offer initial guidelines and rules-of-thumb for selecting appropriate domains under varying channel conditions. We first detail channel expressions and relationships in various domains referring to linear-system models. We then examine signal designs in different domains and disclose their connections. We further review equalization techniques for doubly selective channels, highlight their connections to channel sparsity, and evaluate channel sparsity and equalization performance for four representative channels. A summary and future research opportunities are finally presented for domain-adaptive modulation and equalization, followed by conclusions.

## II. DOMAINS AND WIRELESS CHANNELS

Wireless channels can be characterized by their time-varying and frequency-selective nature. Channel representations in different domains significantly impact system design and analysis. Domains can be defined with respect to (w.r.t.) either signals or channels. This section describes domains and their connections, mainly referring to wireless channels.

Various domains have been defined. Traditionally, we only use two domains: *time and frequency*. To more accurately characterize doubly selective channels, the concepts of *delay* and *Doppler domains* are introduced. The delay domain represents the multipath delay at a specific time, with its Fourier transform counterpart being the frequency domain. The Doppler domain captures the multipath phase variation over time caused by environmental dynamics, and its Fourier transform counterpart is the time domain. The combinations of any two domains can be used to describe an array of channels or signals. In discrete forms, different domains can be linked by one or more DFTs, or inverse DFTs (IDFTs).

There have been three major domains being defined for doubly selective channels: delay-time (dt), frequency-Doppler (fD), and delay-Doppler (dD). We refer to the systems with data modulated in these domains as SC, OFDM, and OTFS, respectively. Consider a system with bandwidth *B*, a block of transmitted signals of length-*P* with a sufficiently long cyclic prefix (CP), and a channel of *L* paths with amplitudes $h_\ell$, and delays $\tau_\ell$ and Doppler frequencies $v_\ell$ that are normalized to *1/B* and *B/P*, respectively, and could have off-grid fractional values. The normalized maximum delay and Doppler spreads are $\mathrm{T_d}$ and $\mathrm{F_d}$, respectively. Similar normalization is used in the axes of figures hereafter. Note that all systems considered in this article have the same signal structure of a P-sample block prepended by a single CP.

Since wireless communication systems are typically linear, we discuss domains with reference to the linear-system signal model and will focus on the channel matrixes derived from the model. Refer to a general signal model $\boldsymbol{y} = \boldsymbol{H}_{ab}\boldsymbol{x}$, where $\boldsymbol{y}$, $\boldsymbol{x}$, and $\boldsymbol{H}_{ab}$ denote the received signal vector, the transmitted one, and the (equivalent) $P \times P$ channel matrix in the domain *ab* between them. Note that $\boldsymbol{y}$ and $\boldsymbol{x}$ do not have to be in the same domain, and $\boldsymbol{x}$ may be the precoded output of the data symbols $\boldsymbol{s}$. The transformative relationship among several main domains of channels is illustrated in Fig. 1. It is noted that in the linear-system model, each dimension of the channel matrix does not always correspond solely to delay, frequency, time, or Doppler. Rather, it often represents a mix of two of them. This will be elaborated when we discuss the delay-Doppler domain and become more evident in Section III.

*Delay-Time Domain* $\boldsymbol{H}_{dt}$*:* This is a 2D representation of the conventional 1D time domain in a linear system. Its *(m,n)*-th matrix element, $\mathrm{m, n} = 0, \cdots, \mathrm{P}-1$, is given by [5, 9]

$$(\boldsymbol{H}_{dt})_{m,n} = G_2\left(\frac{m}{B}\right)\sum_{\ell=1}^{L} h_\ell\, g_1\left(\frac{(m-n-\tau_\ell)_P}{B}\right) e^{\frac{j2\pi m v_\ell}{P}}, (1)$$

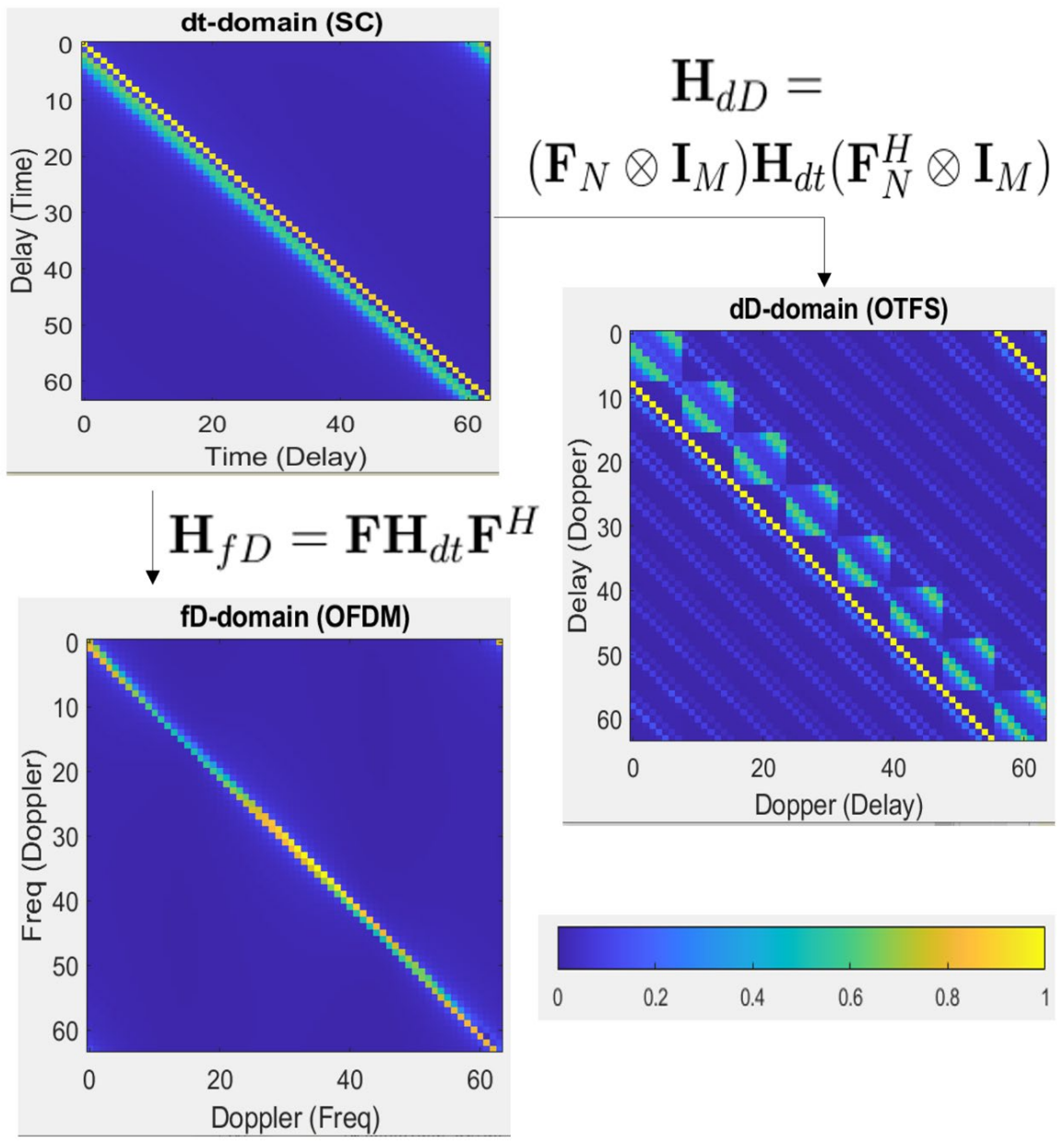

Figure 1 Channel matrices in different domains and their connections in the linear-system model. A small *P* of 64 is used for illustrative clarity. $\mathbf{F}$ and $\mathbf{F}_\mathrm{N}$ denotes *P*- and *N*-point DFT matrices, respectively, $\mathrm{I_M}$ denotes an $M \times M$ identity matrix, $(\cdot)^\mathrm{H}$ denotes conjugate transpose, and $\otimes$ denotes Kronecker product. $\mathrm{M} = N = 8$. Channel setup: path number *L=3*, maximum Doppler spread $\mathrm{F_d} = 1$, maximum delay spread $\mathrm{T_d} = 6$.

where $(\cdot)_P$ denotes modulo-*P*. $\mathrm{G_2}(\cdot)$ and $\mathrm{g_1}(\cdot)$ are the windowing and filtering functions in the time domain, respectively. Without explicitly using windowing and filtering, they represent rectangle and sinc (i.e., the discrete-time Fourier transform, DTFT, of an all-ones sequence) functions, respectively. For time-invariant channels, the dt-domain channel matrix is circulant; however, it is not any more in doubly selective channels where Doppler shifts are nonzero, as can be readily seen from (1). In particular, fractional delay and Doppler cause leak power to neighbouring-grid channel coefficients, which can be characterized by the function $\mathrm{g_1}(\cdot)$. Nevertheless, it is still a *band matrix* with nonzero elements along the diagonals, retaining good sparsity if the delay spread is small.

*Frequency-Doppler Domain* $\boldsymbol{H}_{fD}$: Signals or channels in this domain can be obtained by left-multiplying a DFT matrix and right-multiplying an IDFT matrix to $\boldsymbol{H}_{dt}$. Its *(m,n)*-th element is given by [5, 9]

$$(\boldsymbol{H}_{fD})_{m,n} = G_1\left(\frac{nB}{P}\right)\sum_{\ell=1}^{L} h_\ell g_2\left(\frac{(m-n-v_\ell)_N\, B}{P}\right) e^{-\frac{j2\pi n\tau_\ell}{P}}, (2)$$

where $\mathrm{G_1}(\cdot)$ and $\mathrm{g_2}(\cdot)$ are the DFT and IDFT of $\mathrm{g_1}(\cdot)$ and $\mathrm{G_2}(\cdot)$, respectively. This leads to a sparse band matrix with nonzero elements along several diagonals around the main diagonal. The elements in each row represent frequency channel responses, spread due to Doppler shifts, while elements along each diagonal, i.e., under the same value of *m-n*, have the same Doppler shift. Typically, the main diagonal elements dominate unless the Doppler frequency is larger than $\mathrm{B}/(2\mathrm{P})$.

*Delay-Doppler Domain* $\boldsymbol{H}_{dD}$*:* The dD-domain channel matrix in the linear-system model is different to that in the *original OTFS dD-domain,* although the former can be derived from the latter. The latter is represented by a reduced-size matrix of MxN, MN = P, and the received signal, in an array form, is modelled as a 2D linear quasi-convolution between this channel matrix and the transmitted signals in the array. Rows and columns of the channel matrix well correspond to delay and Doppler, respectively. By vectorizing the signals and transforming this expression into the linear-system model, we obtain the dD-domain channel matrix $\boldsymbol{H}_{dD}$, together with its expression, as shown in Fig. 1 [7]. Each row and column in $\boldsymbol{H}_{dD}$ correspond to mixed delay and Doppler.

It is widely assumed in OTFS literature that channels possess sparsity in the original reduced-size dD-domain; that is, most channel coefficients have negligible power and can be neglected in signal processing. Sparsity is an essential assumption to allow efficient signal processing in the dD-domain. However, we will show such sparsity in $\boldsymbol{H}_{dD}$ is practically occasional, even for simple two-path channels. This is because (1) delay and Doppler values are often off-grid, leading to multiple nonzero channel coefficients, and (2) model linearization leads to further decreased sparsity. Actually, $\boldsymbol{H}_{dD}$ is a diagonal *stripe matrix* with nonzero diagonals spaced at a fixed interval. Its cause will be further explained in Section III.

*Other Domains and Modulating Techniques*: We can also obtain the ft-domain channel matrix by applying a DFT to $\boldsymbol{H}_{dt}$ over the delay domain or the direct dD-domain matrix via applying an IDFT to $\boldsymbol{H}_{fD}$. Neither of them has good channel sparsity. They are less of interest in system design and will be ignored in this article. Meanwhile, new modulating techniques, competing with OTFS, have also been proposed recently [10]. More detailed illustrations of their channels and relationships can also be found in the work.

## III. Signals and their Connections in Different Domains

In this section, we explore and compare signals modulated in various domains, with reference to OFDM and Zak-OTFS, which modulate data symbols in fD- and dD-domains, respectively. By utilizing the *layered inverse Fast Fourier Transform (IFFT) structure* from [11], we can readily establish connections between signals designed in these domains.

These connections are illustrated in Figure 2, which shows a simplified block diagram of the OTFS system interpreted as a precoded OFDM system [11], based on the layered IFFT structure. Using the divide-and-conquer approach, this structure decomposes an IFFT (or FFT) into multiple layers of smaller IFFT (or FFT) operations. For a $P = MN$ point IDFT, it consists of three modules: column-wise $M$-point IDFT, an element-wise phase weighting operation with matrix $\boldsymbol{W}$, and row-wise $N$-point IDFT. Note that the frequency domain vector signals are input to the first module row-wise, and then the third module outputs signals column-wise. OTFS inputs data symbols into the third module, in the dD domain, rather than from the first module in OFDM. It is the same domain as proposed in the asymmetric OFDM system [5], which considers time-invariant frequency selective channels only. The transmitted signal format is also similar to the vector OFDM system [12]. Therefore, referring to the layered IFFT structure, we can interpret OTFS as precoded OFDM with a precoder consisting of a phase weighting operation with element-wise inversion (conjugate) of $\boldsymbol{W}$ and column-wise $M$-point DFT, the inverse of the first two modules in the layered IFFT structure. Such a precoder will fully cancel the first and second modules in the $P$-point IDFT, if the signal order is unchanged in the frequency domain. However, the precoder will become explicit if signal order changes, due to, e.g., adding pilot subcarriers and reserved subcarriers in the frequency domain. Such additions are sometimes necessary in practical systems, as evident from, e.g., the DFT-precoded OFDM in 4G/5G mobile networks.

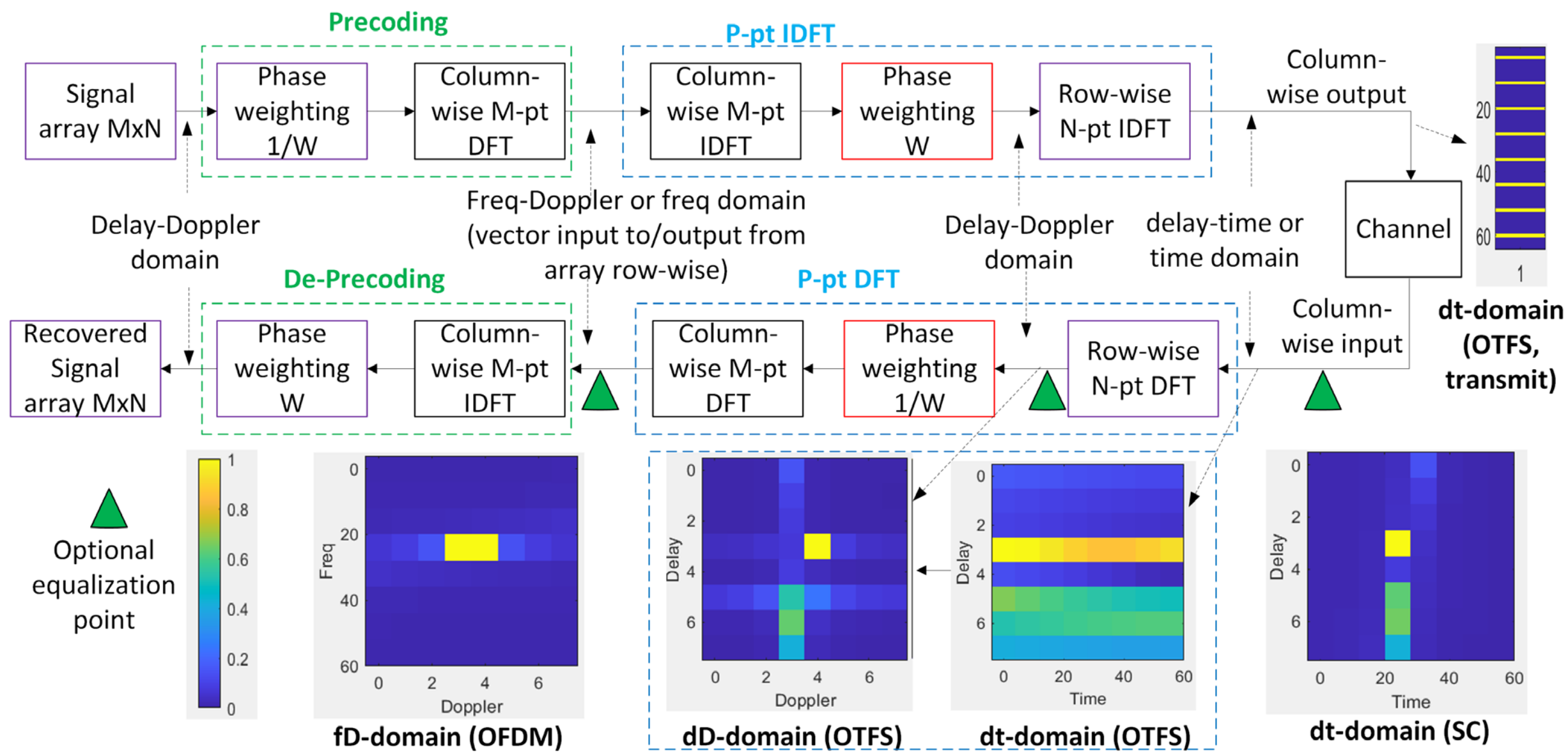

Figure 2 Interpretation of OTFS as precoded OFDM and the received signal patterns. Same channel parameters with those in Fig. 1. The (m,n)-th element of **W** is $\exp(-j2\pi mn/P)$. Additional domains can be defined by modifying the data flow, e.g., replacing "Column-wise output/input" by "Row-wise output/input" before/after Channel, dD-domain will become time-frequency domain.

The OTFS signal formulation process shows that a data symbol spread to $M$ interleaved subcarriers and $N$ interleaved time-domain samples. Hence, OTFS can achieve both frequency and time diversities in doubly selective channels.

Further inspecting receiver processing discloses insights on the dD-domain channel. At the bottom of Fig. 2, some signal patterns, in an MxN array, $M = N = 8$, in different domains for the same channels in Fig. 1 are shown. From left to right, the first and fourth show the received fD-domain and dt-domain signals for a single symbol 1 in the corresponding domain at the transmitter (OFDM and SC modulations, respectively), and the middle two show the dD-domain and dt-domain signals for a single symbol 1 in the dD-domain at the transmitter (OTFS modulation). These signal patterns show how single transmitted symbol spreads in each domain at the receiver. As shown in the block diagram in Fig. 2, the received dt-domain signal vector is reshaped into an MxN array by mapping its elements column-wise. Consider a simple example where $M>L$ and no multipath signal is split between two columns. Refer to OTFS modulation. In the dt-domain, we can see that in the received signal array, each column contains a segment of samples spread from one transmitted dt-domain symbol due to multipath; and each row contains transmitted dt-domain samples weighted by channel coefficients of the same amplitude but time-varying phases due to Doppler shift. Therefore, the dt-domain channel coefficients are re-aligned as per each multipath in each row; and as a result, applying $N$-point DFT row-wise generates a compressed Doppler spectrum of the channel in the dD-domain. However, signals across rows are not compressed in the delay domain, leaving still a relatively large number of nonzero coefficients, as seen from the dD-domain signal pattern.

We can readily link the received signal array here to the channel matrices in the linear-system model. If we vectorize the received array signal column-wise, we will obtain one column of the channel matrix in the linear-system model, as shown in Fig. 1. Thus, we can clearly see how the domains are mixed in each dimension of the linear-model channel matrix. For OTFS, since signals are not compressed in the delay domain, sparsity becomes low when the delay spread is large. This results in observable periodic patterns of non-negligible channel coefficients along the diagonals of the dD-domain channel matrix. These diagonal elements are spaced at intervals of $M$, creating a stripe-structured, repeating pattern in $\boldsymbol{H}_{dD}$, as can be clearly seen from Fig. 1. Comparatively, the fD-domain and dt-domain signals only have a small segment of non-negligible values in one row and column, respectively; therefore, we see their channels in the linear-system model are much sparser.

However, *it is important to note* that, further applying column-wise $M$-point DFT to the dD-domain signal array directly, without using the phase weighting module, does not generate signals of good sparsity. The phase weighting matrix is essential for compressing signals in both delay and time domains, by aligning the phases of signals.

In summary, understanding the signals and channels in different domains and their transformations through structures of the layered IFFT provides valuable insights into the design of advanced wireless communication systems in doubly selective channels. The interpretation of OTFS as a precoded system can also be generalized in several ways.

- First, inspecting the mechanism of achieving frequency and time diversities via signal spreading in OTFS, we see that this functionality can also be realized by other precoders spreading a symbol to multiple domains if desired;
- Second, instead of viewing OTFS as a precoded OFDM system, we can also treat it as a precoded SC system in the dt-domain, with the precoder being the third module in the $P$-point layered IFFT structure;
- Third, the precoder outputs can be shuffled in a different order and/or allocated to partial of the subcarriers in the frequency domain, or partial of the samples in the time domain. This enables flexible resource allocation, pilot design, and out-of-band emission suppression using, e.g., guarding subcarriers.

## IV. Channel Sparsity and Equalization

Interpreting OTFS as a precoded system and generalizing the precoder design, as summarized in the last section, enable us to place signal modulation, pilot insertion and channel estimation, and equalization separately in different domains. This means that they do not have to be in the same domain, and such designs can be adapted to domain-specific channel conditions in various applications. For example, for OTFS, equalization can be in any of the four domains, while the pilot does not have to be in the same domain with either of them.

Factors to be considered for domain selection include *diversity gain, system overhead, processing complexity and equalization performance*. Except for the diversity gain, the other three are closely related to channel sparsity. It is noted that *the Shannon capacity is identical no matter which domain modulated data symbols are placed in*, as the transforms between channels of different domains are all orthonormal. However, equalization can have significantly different performance and complexity in different domains.

Next, we first review typical equalization techniques for doubly selective channels, then examine the sparsity of channels in different domains, and show how system performance may vary with processing in different domains. Since the channel matrix in the ft-domain is not sparse at all, we will only consider dt-, fD-, and dD-domains.

**Equalization Techniques:** We briefly review equalization techniques and comment on their applicability, complexity, and performance in the previously discussed domains.

*Linear equalizers* [5], such as zero-forcing (ZF) and minimum mean square error (MMSE) equalizers, have been widely applied in time-invariant systems. For doubly selective channels, they are also applicable in all the domains mentioned above. Equalization in different domains will perform similarly if the equalizer is constructed from the perfect channel estimate and data symbols are modulated in the same domain. However, equalizers constructed from truncated channel estimates, with pilots designed to save system overhead, may lead to degraded performance. In addition, when the channel matrix is large, equalization complexity becomes prohibitively high for practical implementation. To overcome this issue, various techniques that exploit the channel matrix structure and sparsity have been developed, such as Offset Gradient Descent [6], a specialized algorithm for MMSE, and MMSE simplifying algorithms exploring Hermitian matrix properties [13]. The complexity and performance of such techniques are closely related to the channel sparsity.

*Message passing (MP) algorithms* [1,14], such as belief propagation, are iterative techniques for solving complex estimation and decoding problems. They are introduced for OTFS channel equalization, to exploit the sparsity and address the complicated convolutional relationship between signals and channels in the original dD-domain. It is effective for complex channels with significant inter-symbol interference (ISI), as present in the original dD-domain. However, it has high computational complexity due to its iterative nature, which increases fast with sparsity decreasing, so is its performance [1]. For systems with a loopy graph, such a receiver will have an error floor.

*Interference cancellation equalization,* such as MMSE-serial interference cancellation (SIC) [7] and Turbo equalization [8], exploit the sparse and sub-block structure of channel matrices to demodulate symbols and remove their interference from subsequent sub-block successively. This process repeats until all the symbols are demodulated. The techniques can significantly reduce complexity and can be applied to any domain with a sub-block sparse channel structure. Error propagation is one major challenge, although it can be mitigated via iterative processing. The size and number of the sub-blocks, which impact the complexity and severity of error propagation, depend on the channel sparsity and structure.

The mainstream equalization techniques reviewed above show that their performance and complexity are closely linked to the channel sparsity. Meanwhile, it is noteworthy that the number of pilots required for channel estimation also depends on channel sparsity. *Therefore, channel sparsity shall be a major factor to consider for domain selection.*

**Channel Sparsity**: To evaluate channel sparsity in different domains, we introduce two *power ratio* metrics: *localized-power-ratio (LPR)* and *sorted-power-ratio (SPR)*. The LPR is computed as follows: for the $n$-th column in a $P \times P$ channel matrix, find the element with the peak power and its index $q_n$; then, compute the power ratio between $2L_c + 1$ elements with indexes $(q_n - L_c{:}\, q_n + L_c)_P$ and the total elements; and, finally, compute the ratio average across all $N$ columns. The SPR is similarly computed with the difference that the first $2L_c + 1$ maximal elements in each column are used. These two metrics serve as effective indicators of channel spread and sparsity. The LPR is a better one in terms of the efficiency of implementing equalization, while SPR is beneficial to dD-domain channels, because of band channel matrices in both td- and fD-domains and stripe matrices in the dD-domain.

Four representative channel configurations for different applications are considered, as shown in Table 1. Note that the values of these parameters are only indicative and estimated based on likely physical setups in the scenarios, and they are not exclusive. Channel parameters are randomly generated following uniform distributions between the specified ranges. Channel amplitudes are generated following Rician fading with the Rician factor $R_f$. To be consistent with the OTFS literature, a relatively large *P=1024* (*M=16, N=64)* is used, although this is not necessary because randomly generated (off-grid) delay and Doppler values in the specified ranges are used.

Figure 3 presents the channel LPR and SPR in three domains. It shows that when the delay spread is large (Cases 2 to 4), the fD-domain channel matrix holds the greatest sparsity according to the LPR, due to the channel compression from both delay and time domains. Its LPR and SPR values also match well. The dt-domain channels present good sparsity when the path number is small and the delay spread is not too large (Cases 1 and 2). The LPR for the dD-domain is mostly lowest, even when there are only two paths with small delay spread, due to the scattered nonzero elements of the stripe matrix. The dD-domain SPR is significantly larger than its LPR; however, exploring such scattered power requires very complicated equalizer. In all four cases, either dt-domain or fD-domain channels demonstrate higher sparsity than the dD-domain ones. Note that although only numerical results are presented here, it is possible to analytically characterize the LPR and impact of channel truncation on equalization performance, based on the channel expressions provided in this article.

**Equalization Performance**: We compare the equalization performance for systems under these channels, which are assumed to be perfectly known. Five system setups are considered: SC and OFDM are for those where data symbols and equalization are both in dt- and fD-domains, respectively; For OTFS, OTFS (td Eq) and OTFS (fD Eq), data symbols are all in dD-domain, while equalization is in dD-, td- and fD-domains, respectively. The bit error rates (BERs) for these setups with MMSE equalization are shown in Figure 4. The MMSE equalizers are constructed based on the perfectly known full channel matrix and truncated ones corresponding to the LPR channel truncation. We can see that for full channels, three OTFS setups have the same BER, as expected, outperforming OFDM due to the frequency and time diversity. SC outperforms both OFDM and OTFS because of its full frequency diversity in cases 3 and 4, when Doppler is small and the achievable time diversity is insignificant. For truncated channels, dt-domain equalization achieves significantly lower BER than others when multipath number is small; and fD-domain equalization becomes the best option when multipath number is large. This is because both schemes collect most of the channel power and achieve the highest signal-to-interference ratios. These observations can be broadly applied to most linear equalizers, given their similar operating principles. They can also be extended to advanced equalizers, such as SIC and MP, when operating on a linear system model, as channel truncation errors may dominate the system performance in this case. However, techniques exploring sparsity in other signal-and-channel relationships, such as 2D-quasi-convolution in the original dD-domain, may achieve improved performance at the same channel truncation level but with much higher complexity.

Table 1 Four representative channels cases. $L$, $T_d$ and $F_d$ are as defined in Section II; $R_f$ Rician factor, in dB.

| Case | L | $T_d$ | $F_d$ | $R_f$ | Typical Application Scenarios |
|---|---|---|---|---|---|
| 1 | 2 | 5 | 2 | 10 | LEO satellite |
| 2 | 2 | 8 | 0.5 | 5 | Airplane |
| 3 | 8 | 16 | 0.1 | 6 | High-speed train with a strong LOS path |
| 4 | 8 | 24 | 0.2 | 2 | High-speed train, with a weak LOS path |

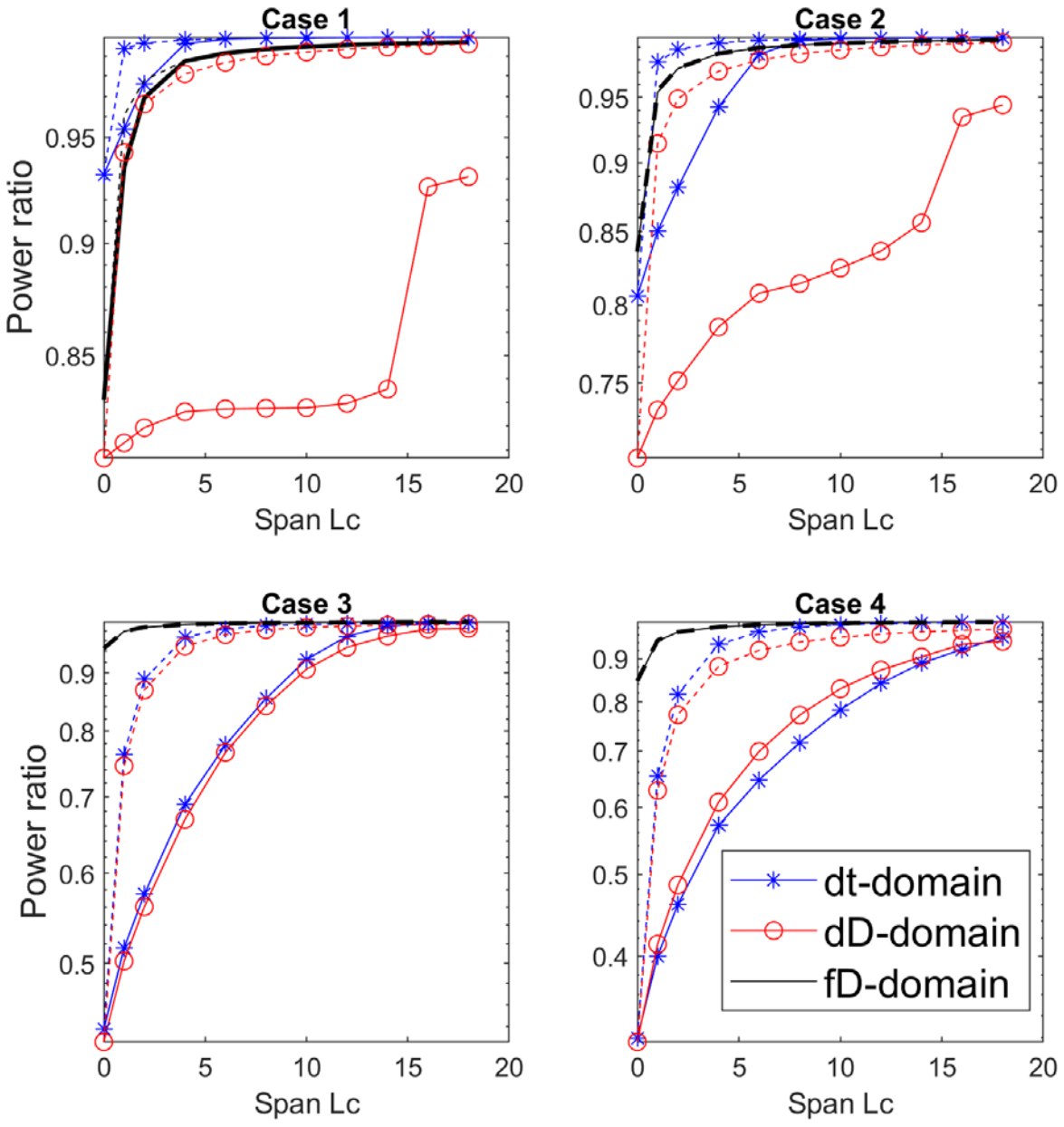

Figure 3 Power ratio of channels in three domains: Solid curves for LPR and dashed curves for SPR.

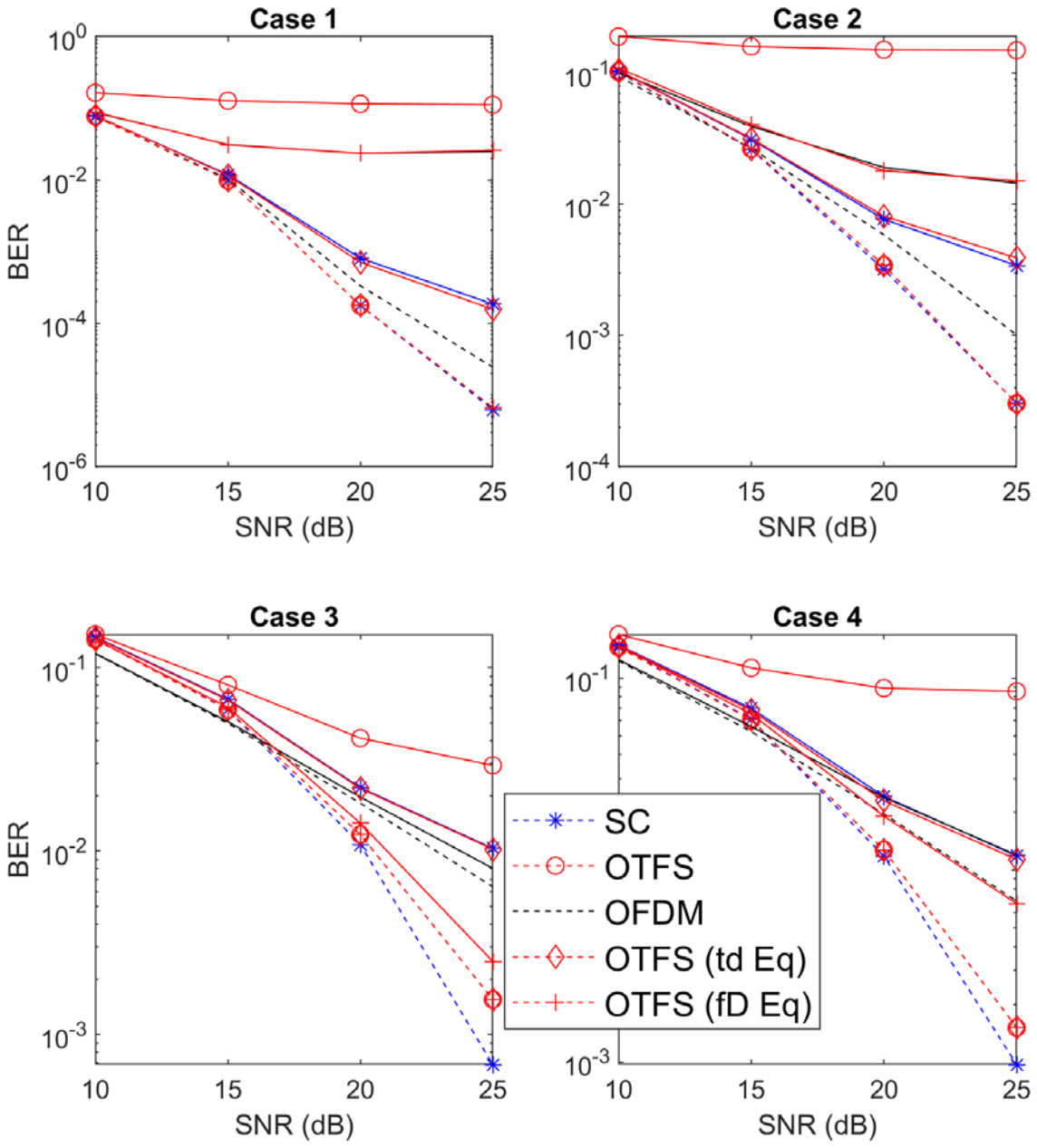

Figure 4 BER of systems with MMSE equalizer in various domains and channels: Dashed and solid curves are for raw and truncated channels, respectively. The truncation parameter $L_c = T_d$ in each case.

## V. Summary of Domain Adaptive Designs and Future Research Opportunities

We summarize the key points for domain adaptive designs:

- Data symbols, pilots, channel estimation, and equalization can be placed in different domains. The modulation domain can be independently considered and designed in relation to the others, with its domain selection being more closely related to the diversity available in the channel. The channel delay and Doppler spreads have a major impact on the diversity, and different orders of diversity can be achieved via adapting, e.g., the values of $M$ and $N$, to the channel conditions. Pilot may be flexibly placed in different domains, as will be elaborated further.
- OTFS can be regarded as a precoded OFDM or SC system. The precoder can also be generalized per the discussions at the end of Section III.
- The best-fit domains for equalization are channel-dependent, as elaborated in the last paragraph of Section IV and summarized in Table 2. While this table lays the groundwork by linking channel conditions to preferred equalization domains, we acknowledge that detailed, quantitative characterization of channels remains an open research challenge.

By leveraging the unique advantages of each domain and employing adaptive techniques, future wireless systems can achieve robust and efficient communication across diverse and challenging environments. Domain adaptive designs open at least the following future major research opportunities.

*Dynamic Domain Adaptation Algorithms*: One of the significant open research problems is developing quantitative criteria for domain adaptation and algorithms that can seamlessly switch between different domains based on channel conditions. We have only demonstrated such a necessity qualitatively. Research can focus on creating quantitative criteria and algorithms based on either channel statistics or continuously monitoring channel characteristics. Such criteria may be based on the power metrics introduced in this article or those associated with the specific equalization techniques. It is also possible to determine the most suitable domains based on real-time channel estimates. The algorithms need to be efficient, minimizing the overhead and latency associated with domain switching, while maximizing the performance benefits.

Table 2 Summary of domain adaptive equalization.

| **Channel Conditions** | **Best-fit Domain for Equalization** |
|---|---|
| Small $T_d$ | dt-domain |
| $T_d$ is large and $F_d$ is not very large | fD-domain |
| Sparse channels in dD-domain | dD-domain |
| Other channels | fD-domain is relatively better |

*Robust Pilot Design and Efficient Channel Estimation Techniques*: Research can focus on developing adaptive pilot placement strategies that can be dynamically adjusted for domains and resource usage based on the current channel conditions. Pilots can be in a different domain with data symbols or equalization, only if their transformed signals in the domain of channel estimation can be well separated from data symbols. Superimposed data and pilot may also be used, when advanced joint channel estimation and equalization techniques are applied. Exploring non-traditional pilot structures, such as orthogonal codes or pseudo-random sequences, may also provide better resilience to mobility-induced impairments.

Channel estimation in doubly selective channels remains a challenging task due to the complex nature of these channels. Innovative approaches such as compressive sensing, joint channel estimation and symbol estimation, and deep learning-based estimation [15] can be explored to improve accuracy and efficiency. Balancing pilot overhead and computational complexity with estimation accuracy will be crucial for practical implementation.

*Low-Complexity Equalization Algorithms*: Another open problem is the development of low-complexity equalization algorithms that maintain high performance in doubly selective channels. Current state-of-the-art techniques like message passing and turbo equalization offer excellent performance but at the cost of high computational complexity. Research can focus on simplifying these algorithms or developing new ones that achieve similar performance with reduced complexity by exploring channel sparsity in various domains. Techniques such as approximations, interference cancellation, and iterative refinement can be investigated to make these algorithms more feasible for real-time applications.

*Integrated Sensing and Communications (ISAC) with Domain Adaptivity*: Future research in ISAC using domain adaptivity presents several exciting opportunities. One area of interest is to investigate the trade-offs between sensing accuracy and communication throughput in different domains and explore the best domains for communications and sensing jointly. It has been shown that the fD-domain is also an excellent option for sensing [9]. Another problem is developing advanced algorithms that dynamically allocate resources across different domains between sensing and communication tasks based on channel conditions.

## VI. Conclusion

We have explored different domains and their connections for doubly selective channels, evaluated the channel sparsities in these domains, and illustrated their impact on system performance. It is shown that the selected domain, particularly for equalization, significantly affects system performance. By allowing modulation, pilot insertion and equalization in different domains and making equalization domain-adaptive, we can unlock new opportunities and boost performance and efficiency for communications in doubly selective channels.

## References

1. P. Raviteja, K. T. Phan, Y. Hong and E. Viterbo, "Interference Cancellation and Iterative Detection for Orthogonal Time Frequency Space Modulation," in IEEE Trans. on Wireless Communications, vol. 17, no. 10, pp. 6501-6515, Oct. 2018
2. Z. Wei et al., "Orthogonal Time-Frequency Space Modulation: A Promising Next-Generation Waveform," in IEEE Wireless Communications, vol. 28, no. 4, pp. 136-144, August 2021
3. H. Lin and J. Yuan, "Orthogonal Delay-Doppler Division Multiplexing Modulation," in IEEE Transactions on Wireless Communications, vol. 21, no. 12, pp. 11024-11037, Dec. 2022
4. S. K. Mohammed, R. Hadani, A. Chockalingam and R. Calderbank, "OTFS - Predictability in the Delay- Doppler Domain and Its Value to Communication and Radar Sensing," in IEEE BITS the Information Theory Magazine
5. H. Zhang, X. Huang and J. A. Zhang, "Adaptive Transmission With Frequency-Domain Precoding and Linear Equalization Over Fast Fading Channels," in IEEE Trans. on Wireless Communications, vol. 20, no. 11, pp. 7420-7430, Nov. 2021
6. H. Zhang, X. Huang and J. A. Zhang, "Fine Doppler Resolution Channel Estimation and Offset Gradient Descent Equalization for OTFS Transmission over Doubly Selective Channels," 2023 22nd ISCIT, Sydney, Australia, 2023
7. Z. Wang, Z. Liu, F. Xing, R. Sun and X. Ning, "Low Complexity MMSE-SIC Receiver for OTFS in High-Speed Mobile Scenarios," in IEEE Communications Letters, vol. 28, no. 3, pp. 667-671, March 2024
8. Q. Li, J. Yuan, M. Qiu, S. Li and Y. Xie, "Low Complexity Turbo SIC-MMSE Detection for Orthogonal Time Frequency Space Modulation," in IEEE Trans. on Communications, vol. 72, no. 6, pp. 3169-3183, June 2024
9. Y. Sun, J. A. Zhang, K. Wu and R. P. Liu, "Frequency-Domain Sensing in Time-Varying Channels," in IEEE Wireless Communications Letters, vol. 12, no. 1, pp. 16-20, Jan. 2023
10. Hyeon Seok Rou, et al., “From OTFS to AFDM: A Comparative Study of Next-Generation Waveforms for ISAC in Doubly-Dispersive Channels”, arXiv:2401.07700, 13/10/2024
11. J. Zhang, A. D. S. Jayalath and Y. Chen, "Asymmetric OFDM Systems Based on Layered FFT Structure," in IEEE Signal Processing Letters, vol. 14, no. 11, pp. 812-815, Nov. 2007
12. X. Xia, "Precoded and vector OFDM robust to channel spectral nulls and with reduced cyclic prefix length in single transmit antenna systems," in IEEE Trans. on Communications, vol. 49, no. 8, pp. 1363-1374, Aug. 2001
13. G. D. Surabhi and A. Chockalingam, "Low-Complexity Linear Equalization for OTFS Modulation," in IEEE Communications Letters, vol. 24, no. 2, pp. 330-334, Feb. 2020
14. W. Yuan, Z. Wei, J. Yuan and D. W. K. Ng, "A simple variational Bayes detector for orthogonal time frequency space (OTFS) modulation", IEEE Trans. Veh. Technol., vol. 69, no. 7, pp. 7976-7980, Jul. 2020.
15. Z. Zhou, L. Liu, J. Xu and R. Calderbank, "Learning to Equalize OTFS," in IEEE Trans. on Wireless Communications, vol. 21, no. 9, pp. 7723-7736, Sept. 2024